\documentclass[manuscript,preprint,epsfig,amsmath]{aastex}

\newcommand{\be}{\begin{equation}}
\newcommand{\ee}{\end{equation}}
\newcommand{\bea}{\begin{eqnarray}}
\newcommand{\eea}{\end{eqnarray}}

\shorttitle{Hot Jupiter Trojans}
\shortauthors{Ford and Gaudi}
\begin{document}

\title{Observational Constraints on Trojans of Transiting Extrasolar Planets}
\author{Eric B.\ Ford\altaffilmark{1,2} and B.\ Scott Gaudi\altaffilmark{2,3}}

\email{eford,sgaudi@cfa.harvard.edu}
\altaffiltext{1}{Hubble Fellow}
%\altaffiltext{2}{Center for Astrophysics, Mail Stop 51, 60 Garden Street, Cambridge, MA 02138}
\altaffiltext{2}{CfA, MS 51, 60 Garden St., Cambridge, MA 02138}
\altaffiltext{3}{Current address: Department of Astronomy, The Ohio State University, 140 W.\ 18th Ave., Columbus, OH 43210}

\begin{abstract}
Theoretical studies predict that Trojans are likely a frequent
byproduct of planet formation and evolution.  We present a novel
method of detecting Trojan companions to transiting extrasolar planets
which involves comparing the midtime of eclipse with the time of
the stellar reflex velocity null.  We demonstrate that this method
offers the potential to detect terrestrial-mass Trojans using existing
ground-based observatories.  This method rules out Trojan companions
to HD 209458b and HD 149026b more massive than $\simeq~13~M_{\oplus}$
and $\simeq~25~M_{\oplus}$ at a 99.9\% confidence level.  Such a
Trojan would be dynamically stable, would not yet have been detected
by photometric or spectroscopic monitoring, and would be
unrecognizable from radial velocity observations alone.  We outline
the future prospects for this method, and show that the detection of a
``Hot Trojan'' of any mass would place a significant constraint on
theories of orbital migration.
\end{abstract}
\section{Introduction} 
Stable Trojan companions to extrasolar planets may be common.  In our
solar system, Mars, Jupiter, and Neptune share their orbit with
asteroids orbiting near the stable (L4/L5) Lagrange points that
lead/trail the planet by $\simeq~60^{\circ}$.  Orbits near the L4/L5
points of the terrestrial planets, Saturn and Uranus are 
less stable due to perturbations from the other planets (Nesvorny \&
Dones 2002).  Saturn's satellites also include small moons
orbiting about the L4/L5 points of Tethys and Dione.  While the mass
ratios of the Trojan systems in our solar system are rather extreme ($\le7\times10^{-9}$),
extrasolar planets may have more massive
Trojans.  Theorists have already outlined several
mechanisms to form Trojans with mass ratios
as large as unity.  For example, Laughlin \& Chambers (2002)
present hydrodynamic simulations of a
protoplanetary disk where disk material lingers near the
L4 and L5 points of a planet (near the gap-opening threshold).
The resulting vortex could trap particles and lead to the accretion of
a Trojan {\em in situ} (Chiang \& Lithwick 2005).  If disk torques 
caused the planet to gradually migrate inwards, then the Trojan would
migrate with the planet.  Unlike resonant migration in the 2:1 mean
motion resonance, the eccentricity or libration amplitude of the
Trojan would not be excited by the migration (Laughlin \& Chambers
2002).
Alternatively, a body could be captured into an orbit about the L4/L5
point after a violent event, as has been suggested for the formation
of Jupiter and Neptune Trojans in our solar system (Morbidelli et al.\
2005).  Capture into a Trojan orbit could also occur due to rapid mass
growth of the planet or a collision of two objects near L4/L5 (Chiang
\& Lithwick 2005 and reference therein).  Another possibility is
that convergent migration could trap multiple protoplanets into
a 1:1 mean motion resonance (Thommes 2005, Cresswell \& Nelson 2006).  
In each of these
scenarios, the captured bodies could initially have a large libration
amplitude or reside on horseshoe-type orbits.  However, if the capture
occurred before or during the planet's inward
migration, then interactions with either a gaseous or planetesimal
disk would damp the libration amplitude.  This
mechanism is even capable of causing objects initially on horseshoe
orbits to evolve into tadpole orbits and then small amplitude
libration near the L4/L5 fixed point.  Such behavior has been found
in numerical simulations of multiple planet systems interacting
with either a gaseous or planetesimal disk (Cresswell \& Nelson 2006; Ford \& Chiang 2006). 

Trojans of Jupiter and Neptune have
provided clues about our solar system's history
(Michtchenko, Beauge \& Roig 2001; Kortenkamp, Malhotra \& Michtchenko
2004; Chiang \& Lithwick 2005; Morbidelli et al.\ 2005).  Similarly,
the detection of extrasolar Trojans would be useful for
constraining theories of planet formation.  
While all the above mechanisms predict that Trojans would survive the
migration process, there are alternative models of planet migration
that predict Trojans would not survive.  
For example,
while planet formation models generally agree that planets should form
on nearly circular orbits, it is possible that gravitational
perturbations by other planets or a binary companion could excite
sizable eccentricities.  One possible formation mechanism
for short-period giant planets is that a planet acquires a large
eccentricity (e.g., due to strong planet-planet scattering, secular
perturbations from a binary companion, or being tidal captured) and
comes so close to the star that tidal dissipation 
circularizes the orbit at a small semimajor axis
(Rasio \& Ford 1996, Wu \& Murray 2003, Gaudi 2003,
Ford \& Rasio 2006).  Detecting
a Trojan companion to a short-period planet would present a serious
challenge for these mechanisms for forming ``hot Jupiters'' and would
imply that the planet in such a system was formed via migration
through a dissipative disk.  
Thus, searching for
extrasolar Trojans can test models of planet formation.
%
%
%
%This test is very
%complementary to observations of the Rossiter-McLaughlin effect that
%can measure the inclination of a planet's orbital plane to the
%rotation axis of the star (e.g.\ Winn et al.\ 2005).  If observations of the
%Rossiter-McLaughlin effect implied a large inclination for a
%short-period planet, then that would challenge models of migration via
%interactions with a dissipative disk and would suggest formation
%via the excitation of a large eccentricity and
%inclination followed by a phase of tidal circularization (Gaudi \& Winn 2006).  
%Thus, both measurements of the Rossiter-McLaughlin effect and searching for
%Trojan companions of short period planet can provide significant and
%complementary constraints on models of planet formation.
%
%
Here, we present a method for detecting Trojan companions to
extrasolar planets by combining RV and photometric observations of
transiting extrasolar planets.  We refer to all bodies librating about
the L4/L5 fixed point of a planet as ``Trojans'' and 
focus our attention on Trojans that are
significantly less massive than the currently known planet and
not currently recognizable from radial velocity (RV) observations alone.
\section{Observational Constraints on Trojans}
We denote the stellar mass ($m_{\star}$), the planet mass ($m_p$), and
the Trojan mass ($m_T$).  Since the known transiting planets have
short orbital periods and are subject to rapid eccentricity damping
(Rasio et al.\ 1996), we initially assume the planet to be on
a circular orbit about a star.  Then
a Trojan would orbit at one of the two fixed
points, L4/L5, which lie along the orbit of the planet and lead/trail
the planet by $60^{\circ}$.  If there are no other bodies in
the system, then the L4/L5 fixed points are stable if the ratio,
$\mu=(m_p+m_T)/(m_\star+m_p+m_T)$, is less than a critical threshold
$\mu_c$, where $0.03812\le\mu_{c}\le0.03852$ and $\mu_c$ depends on
$\epsilon\equiv m_T/(m_p+m_T)$ (Murray \& Dermott 2000).
If the Trojan resides exactly at the L4/L5 fixed point, then
the direction of the vector sum
of the forces exerted on the star by the planet and Trojan will
lead/trail the force exerted on the star by the planet alone by an
angle, $\phi$, such that
$\tan\phi\simeq~\sqrt{3}\epsilon/(2-\epsilon)\times(1+O(\mu))$ (Fig.~1).

More generally, for a Trojan that is librating about the L4/L5 fixed point, $\phi$ will vary by
an angle $\left|\Delta\phi\right|\sim~\Delta\phi_{\mathrm{fast}}+\Delta~\phi_{\mathrm{slow}}$, where $\Delta\phi_{\mathrm{fast}}$
varies on the orbital period of the planet $P$, and $\Delta
\phi_{\mathrm{slow}}$ varies on the secular
timescale, $P_{lib}\simeq~P\sqrt{4/27}\mu^{-1/2}$ (Murray \& Dermott 2000).
Since the Trojans of short period planets are likely to have formed
in the presence of a dissipative disk,
we focus on Trojans undergoing small
librations with an amplitude, $\delta a \ll \mu^{1/2} a_p$,
where $a_p$ is the semimajor axis of the planet.  
In this case,
$\Delta \phi_{\mathrm{fast}}\sim~e_T \epsilon$, where $e_T$
is the Trojan's osculating eccentricity, and  $\Delta
\phi_{\mathrm{slow}}\sim~[\delta~a/(6 \mu^{1/2}a)]\epsilon$. 
Nesvorny et al.\ (2002) show that the behavior of
Trojans of planets with small eccentricities is
similar.  Perturbations (e.g., GR, stellar quadrupole, tides) are unlikely to affect the stability, since they typically have timescales much longer than the orbital or libration timescale.

If a planet on a circular orbit were the only body perturbing the
central star, then the time that the stellar RV equals the RV of the
system barycenter ($T_0$) would coincide with the time of
midtransit ($T_c$).  However, the gravitational
perturbation of a Trojan at L4/L5 would cause these two times to differ by
\bea
\label{EqnDt}
\Delta t & \equiv & T_0 - T_c = \pm \frac{\phi P}{2\pi} 
                    \simeq \pm\frac{\sqrt{3}\epsilon P}{4\pi} \\
& \simeq & \pm37.5\left(\frac{P}{3\mathrm{d}}\right) \left(\frac{m_T}{10 m_{\oplus}}\right) \left(\frac{0.5 M_J}{m_p+m_T}\right) \mathrm{min}. \nonumber 
\eea
A Trojan can signal its presence by a time
offset between the ephemeris determined from transit photometry and
the ephemeris determined from RVs (Fig.~1).
%  
% ADDED
%
If there are Trojans at both L4 and L5, then this measures the difference in mass at L4 and the mass at L5.
If a planet is on a slightly eccentric orbit, then there is an offset
of $\Delta~t\simeq~P/(2\pi)\times(e\cos\omega+O(e^2))$, where $\omega$ is the argument of pericenter, even in the
absence of a Trojan.  While short-period planets are
expected to circularize rapidly, it is desirable to constrain the
eccentricity observationally (e.g., by RV observations
or timing of the secondary eclipse) before claiming the
detection of a Trojan.  
Additional planets could also perturb the time of
midtransit (Holman \& Murray 2005; Agol et al.\ 2005) so
the offset will vary from
transit to transit.  Therefore multiple transits should be observed to
verify that any observed offsets are not due to perturbations by a
more distant planet.

For a transiting planet, both  $P$ and $T_c$ can be measured
precisely using photometry alone.  Consider a series of continuous photometric
observations with uncorrelated Gaussian uncertainties of magnitude $\sigma_{ph}$,
taken at a rate $\Gamma$ around a single transit. The 
transit time can be measured with an accuracy of
$\sigma_{T_c}\simeq~\sqrt{t_e/2\Gamma}\sigma_{ph}\rho^{-2}$, where
$t_e$ is the duration of ingress/egress
and $\rho$ is the ratio of the planet radius to stellar
radius.  For typical parameters (e.g., $\sigma_{ph}\sim~10^{-3}$), 
$T_c$ can be measured to $\simeq10$s (e.g.,
Brown et al.\ 2001).  The period can be measured
much more accurately, from observations of multiple
transits separated by many orbits.

Given the measurement precision
for $T_c$ and $P$, the practical limit on measuring $\Delta t$ is
set by the
uncertainty in $T_0$ from RV observations.  
Assuming a circular orbit, the RV observations at a time $t_i$ can be
fit by the model\footnote{Ignoring any observations during primary transit when the
Rossiter-McLaughlin effect distorts the observed RV (Winn et al.\
2005; Gaudi \& Winn 2006).} $v_i
= C + K \sin\left(2\pi(t_i-T_c)/P+\phi \right) = C +
A\sin\left(2\pi(t_i-T_c)/P\right) + B\cos\left(2\pi(t_i-T_c)/P\right)$.
where $A \equiv K\cos{\phi}$, $B\equiv K\sin{\phi}$, and $K$ is the velocity semi-amplitude.
Assuming the period determined from photometric
observations, the coefficients $A$, $B$, and $C$, (and hence the phase
difference, $\tan \phi = B/A$) can be determined by linear least squares fitting to RV
observations.  If there
are $N_{RV}$ RV observations with
uncorrelated Gaussian uncertainties ($\sigma_{RV}$)
and many RV observations are evenly distributed over
orbital phase, then a Fisher information analysis (Gaudi \& Winn 2006) reveals that the
uncertainties in model parameters will approach $\sigma_{A} =
\sigma_{B} = \sqrt{2/N} \sigma_{RV}$,
$\sigma_{\phi} = \sqrt{2/N_{RV}}
\sigma_{RV}/K$, and $\sigma_{\Delta t}\simeq~\sigma_{T_0} = 
\sqrt{1/2\pi^2N_{RV}} P \sigma_{RV}/K$.  
A similar analysis for an eccentric orbit in the
epicyclic approximation,
shows that the uncertainty in $\Delta t$ is increased by a modest
factor over the expression above. 
If a Trojan were present, then the uncertainty in
$\phi$ would set the uncertainty in the measurement of the
mass of the Trojan,
\bea
\label{EqnMinMass}
\sigma_{m_T} & = & \frac{4\pi}{\sqrt{3}} m_P \frac{\sigma_{\Delta t}}{P} = \sqrt{\frac{8}{3N_{RV}}} m_P \frac{\sigma_{RV}}{K} \\
& = & 0.52 M_{\oplus} \left(\frac{50}{N_{RV}}\right)^{1/2} \left(\frac{\sigma_{RV}}{\mathrm{m}\, \mathrm{s}^{-1}}\right) \left(\frac{P}{3\mathrm{d}}\right)^{1/3} \left(\frac{m_{\star}}{M_{\odot}}\right)^{2/3}. \nonumber
\eea
If we were to demand a measurement of $\Delta
t>3.291\sigma_{\Delta t}$ to claim the detection of a Trojan,
then a total of $\simeq~160$
(60) precision RV measurements could detect a $\simeq~3M_{\oplus}$
(5$M_{\oplus}$) Trojan, assuming a host star with an
intrinsic jitter, $\sigma_j\simeq 3~{\rm m~s}^{-1}$ (Wright 2005), and
1~m~s$^{-1}$ measurement uncertainties added in
quadrature.   While challenging, it is remarkable that
current ground based instruments have the necessary precision to
detect such a low mass Trojan with a plausible amount of observing
time. 

\section{Example Application}
In Table 1, we summarize the current observational parameters and
sensitivity to Trojan companions of extrasolar planets that
transit bright stars, based on the above analysis.
We find that combining the above method with existing
observations already provides significant upper limits on the mass of
Trojan companions to the planets HD 209458b and HD 149026b.
Next, we perform more careful
Bayesian analyses of the current observational constraints for these two
cases.
For HD 209458b we adopt the transit period and ephemeris 
of Knutson et al.\ (2006). We
reanalyzed the RV measurements from
Butler et al.\ (2006), fixing the orbital period
and transit ephemeris.  We use Markov chain Monte Carlo (Ford 2005, 2006; Gregory 2005)
to sample from the posterior probability distribution for the remaining
RV model parameters $K$, $e$, $\omega$, $M_0$, $C$, and $\sigma_j$,
where $M_0$ is the mean anomaly at the epoch of midtransit.
We assume priors that
are flat in $\log (1+K/K_o)$, $e$, $\omega$, $M_0$, $C$, and
$\log(1+\sigma_j/\sigma_o)$, and choose $K_o=\sigma_o=1$~m~s$^{-1}$, but our
results are insensitive to these assumptions.  We
then construct the posterior distribution for the quantity
$(M_0-e\cos\omega)P/(2\pi) \simeq \Delta t$. 
In Fig.~2a we show the distributions for $\Delta t$ using three different assumptions. 
We find $\Delta t=-11.4\pm8.7$~min (circular orbit), $\Delta t=-16.4\pm10.8$~min (eccentric orbit ignoring the
secondary eclipse), and $\Delta t=-13.1\pm8.9$~min (eccentric orbit using the time of the 
secondary eclipse; Deming et al.\ 2005). We conclude that existing observations place an upper limit on the mass
of Trojan companions to HD 209458b of $13.2M_{\oplus}$ at the 99.9\% confidence level.

We have performed a similar analysis of HD 149026b (Fig.~2b) using the 
observations of Butler et al.\ (2006) and Charbonneau et al.\ (2006).
If we assume a circular (eccentric) orbit, then we find $\Delta t=-19\pm31$~min ($\Delta t=98\pm112$~min). The
constraint is significantly weaker when we allow for an eccentric
orbit, due to the limited number of RV observations and
poor phase coverage.
Incorporating a preliminary estimate of the time of the secondary eclipse (J. Harrington 2006, private communication), we find $\Delta~t=13\pm27$~min and place an upper limit
on the mass of Trojan companions to HD 149026b of $24.5M_{\oplus}$
at the 99.9\% confidence level.

\section{Discussion}
In principle, Trojans could be detected via their radial
velocity, astrometric, transit, or transit timing signatures.  
If a Trojan is sufficiently massive and has a sufficiently
large libration amplitude, then it could be detected from the
deviations from a Keplerian perturbation to the stellar radial
velocity or astrometric signal caused by a single planet.  Laughlin \&
Chambers (2002) have shown that two comparable mass planets occupying
a 1:1 mean motion resonance would typically
have strong planet-planet gravitational interactions on a secular timescale.
However, these signatures may not be unique: a
reanalysis of the RV observations of HD 128311 and HD
82943 have shown that both of the current data sets are consistent
with a pair of planets 
in a 1:1
mean motion resonance (Gozdziewski \& Konacki 2006), as well as the
originally published orbital solutions involving higher-order mean
motion resonances.

Trojans may also be detectable if they transit
their parent star.  
Photometric or spectroscopic
monitoring of stars with transiting planets 
(particularly at times offset from the planet transit by $\sim~P$/6) may reveal the Trojan transit via the decrease in
stellar flux or anomalous RV excursions due to
the Rossiter-McLaughlin effect (Gaudi \& Winn 2006).  For planets discovered via a photometric transit search, there will typically be observations at epochs useful for searching for large Trojans.  Unfortunately, 
a Trojan might not transit its parent star if it has a significant inclination (e.g., Morbidelli et al.~ 2005).  Since the
libration period can be large, long-term monitoring would be
required to ensure detection.
Currently, the most stringent photometric constraints on
Trojan companions to HD 209458b come from the continuous
photometry of the system for 14 days by the MOST
satellite (Rowe et al.\ 2006).   When heavily binned into $\simeq~2$~hr intervals,
this photometry has a fractional uncertainty of $\simeq~3\times10^{-4}$ (Rowe et al.\ 2006).
Assuming an average density equal to that of Earth, this corresponds
to a 3-$\sigma$ detection threshold of $48M_{\oplus}$.  It is not
clear whether the data reduction techniques used in their analysis of
HD 209458 might subtract part of the signal due to a Trojan (Rowe et
al.\ 2006).  Regardless, if a Trojan had a vertical libration
amplitude greater than $\simeq~9^{\circ}$,
it would not always transit the star.  Since the putative libration period
of $\simeq 53$~d (Murray \& Dermott 2000) is significantly longer than the duration of the MOST observations,
it could have missed even a much larger Trojan.  

In a sense, our method is most similar and complementary to the
recently-proposed method of searching for gravitational perturbations due
to low mass planets using transit timing (Holman \& Murray 2005; Agol
et al.\ 2005).  In contrast to the transit timing method, the unique
geometry of Trojan orbits results in a nearly constant perturbation
(assuming small amplitude libration about L4/L5) that recurs at {\em
every} transit.  In principle, it is not necessary to make
precise measurements of the time of many transits to search
for a complex pattern of perturbations.  Thus, our method can be practically applied to transiting
planets with long periods.  Nevertheless, multiple transits should be observed to
ensure that the same offset is observed and avoid
potential confusion with perturbations from a more distant planet
(Holman \& Murray 2005; Agol et al.\ 2005).

In principle, our technique could be applied to search for terrestrial-mass
Trojans of giant planets orbiting in the habitable zone of their stars
(Ji et al.\ 2005; Schwarz et al.\ 2005).  While present search techniques are strongly
biased towards finding transiting planets at short orbital periods,
future space missions (e.g., Corot, Kepler) offer the prospect of
finding transiting planets in the habitable zone of their stars,
particularly for low mass stars where the habitable zone can be 
$\simeq~0.015$~AU away from the star.

\acknowledgments
We thank E. Agol, T. Beatty, E. Chiang, D. Fabrycky, M. Holman, R. Nelson, and J. Winn for helpful comments.  We thank Joe Harrington for providing the time of secondary eclipse for HD 149026b.  
Support for E.B.F.\ was provided by a Miller Research Fellowship
and by NASA through Hubble Fellowship grant
HST-HF-01195.01A awarded by the Space Telescope Science Institute,
which is operated by the Association of Universities for Research in
Astronomy, Inc., for NASA, under contract NAS 5-26555.  Support
for B.S.G.\ was provided by a Menzel Fellowship from 
Harvard College Observatory.
\begin{deluxetable}{lccccccccl}
\tablecaption{Sensitivity to Trojans of Extrasolar Planets}
% that Transit Bright Stars}
\tablehead{
\colhead{Star} &
\colhead{P} &
\colhead{K} &
\colhead{$\sigma_{RV}$\tablenotemark{a}} &
\colhead{$N_{RV}$} &
\colhead{$M_{\star}$} &
\colhead{$\sigma_{\Delta t}$\tablenotemark{b}} &
\colhead{$\sigma_{m_{T}}$\tablenotemark{b}} &
\colhead{References} \\ 
\colhead{} &
\colhead{(d)} &
\colhead{(m s$^{-1}$)} &
\colhead{(m s$^{-1}$)} &
\colhead{} &
\colhead{($M_{\odot}$)} &
\colhead{(min)} &
\colhead{($M_{\oplus}$)} &
\colhead{}
}
\startdata
HD 209458 & 3.5247455(2) & 84(1) & 5.0 & 64 & 1.101 & 8.5 &  2.6 &  1,2   \\
HD 149026 & 2.87598(1)  &  43(2) & 5.7 & 16 & 1.30 & 30.9 &  6.2 &  1,3,4,5 \\
HAT-P-1  & 4.46529(9)  &  60(2) & 5.1 & 13 & 1.11 & 34.0 &  6.4 &  6     \\
TrES-2    & 2.47063(1)  & 181(3) & 6.9 & 11 & 1.08 &  9.2 &  7.6 &  7    \\
HD 189733 & 2.21857(2)  & 205(6) & 15  & 24 & 0.82 & 10.7 &  8.9 &  8,9   \\
TrES-1    & 3.030065(8) & 115(6) & 14  &  8 & 0.89 & 42.3 & 17.  &  1,10   \\
XO-1      & 3.94153(3)  & 116(9) & 15  &  6  & 1.0 & 67.4 & 25.  &  11,12     \\

\enddata
\tablenotetext{a}{When available, we list the rms velocity to the
published best-fit RV model rather than the quoted measurement
uncertainty.}
\tablenotetext{b}{We list ``1-$\sigma$'' uncertainties, implicitly assuming
  circular orbits for the transiting planets.}
\tablerefs{(1) Butler et al.\ 2006; 
(2) Knutson et al.\ 2006; 
(3) Charbonneau et al.\ 2006; 
(4) Sato et al.\ 2005; 
(5) Harrington et al.\ 2006;
(6) Bakos et al.\ 2006b; 
(7) O'Donovan et al.\ 2006; 
(8) Bouchy et al.\ 2005; 
(9) Bakos et al.\ 2006a; 
(10) Alonso et al.\ 2004; 
(11) McCullough et al.\ 2006; 
(12) Holman et al.\ 2006
}
\end{deluxetable} 
\begin{figure}[htbp]
\epsscale{0.7}
\plotone{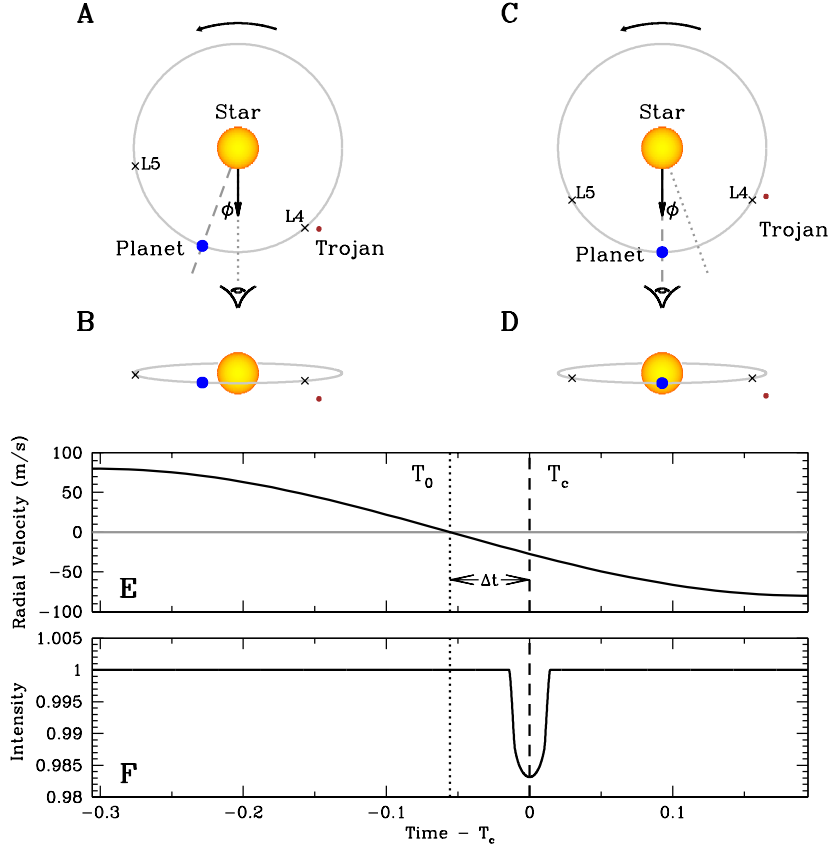}
\caption[fig1]{
%
%\hbox{\plotone{f1.eps} }  \noindent Fig.\ 1:
%
Illustration of the method to detect Trojan companions of 
transiting planets by comparing the transit and RV
observations.  Views of the star, planet and Trojan ({\bf A,C} plan, {\bf B,D} from observer's perspective; not to scale).
The grey circle shows the orbit of the planet and Trojan. The dotted
line indicates the direction of the acceleration of the star, the dashed
line the direction of the transiting planet, and $\phi$ is the angle between these
two directions.  The vector shows the direction toward
the observer.  Panels ({\bf A,B}) show the position at $T_0$, the time of the stellar reflex RV
null. Panels ({\bf C,D}) show the position at $T_c$, the time of the central transit.
Panel ({\bf E}) shows the stellar reflex RV as a function of time  (in units of the period of the planet),
with the times $T_0$ and $T_c$ indicated. Panel ({\bf F}) shows the 
intensity of the star as a function of time.  
We have assumed that the Trojan is inclined so that it does not transit the parent star.  
\label{Fig1}}
\end{figure}
\begin{figure}[htbp]
\plotone{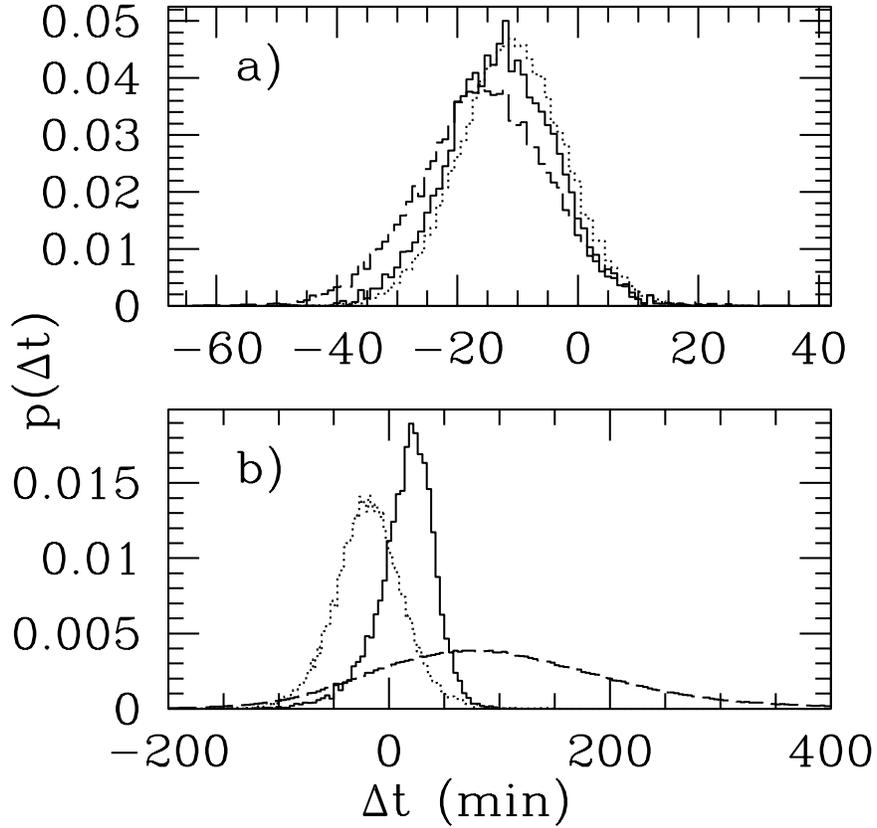}
\caption[fig2]{
%
%\hbox{\plotone{f1.eps} }  \noindent Fig.\ 1:
%
Marginal posterior probability distributions of $\Delta~t$ for HD 209458b (top)
and HD 149026b (bottom).  Here $\Delta t\simeq~(M_0-e\cos\omega)P/(2\pi)$ is the difference
between the time of the stellar reflex RV null 
and the time of central transit
that could be due to a Trojan.  The dotted curves assume a circular
orbit, the dashed curves allow for a non-circular orbit (ignoring the
constraint from the secondary eclipse), and the solid curves allow for
a non-circular orbit and incorporates the measured times of secondary
eclipse.  
\label{Fig2}}
\end{figure}


\begin{thebibliography}{}
%

\bibitem[Agol et al.(2005)]{2005MNRAS.359..567A} Agol, E., Steffen, J., 
Sari, R., \& Clarkson, W.\ 2005, \mnras, 359, 567 

\bibitem[Alonso et al.(2004)]{2004ApJ...613L.153A} Alonso, R., et al.\ 
2004, \apjl, 613, L153 

\bibitem[Bakos et al.(2006)]{2006astro.ph..3291B} Bakos, G.~A., et al.\ 
 2006a, preprint (astro-ph/0603291)

\bibitem[Bakos et al.(2006)]{2006astro.ph..9369B} Bakos, G.~A., et al.\ 
 2006b, preprint (astro-ph/0609369)

\bibitem[Bouchy et al.(2005)]{2005A&A...444L..15B} Bouchy, F., et al.\ 
2005, \aap, 444, L15

% \bibitem[Brasser \& Lehto(2002)]{2002MNRAS.334..241B} Brasser, R., \& Lehto, H.~J.\ 2002, \mnras, 334, 241 


\bibitem[Brown et al.(2001)]{2001ApJ...552..699B} Brown, T.~M., 
Charbonneau, D., Gilliland, R.~L., Noyes, R.~W., \& Burrows, A.\ 2001, 
\apj, 552, 699 

\bibitem[Butler et al.(2006)]{2006ApJ...646..505B} Butler, R.~P., et al.\ 
2006, \apj, 646, 505 

%\bibitem[Charbonneau et al.(2006)]{2006astro.ph..3376C} Charbonneau, D., 
%Brown, T.~M., Burrows, A., \& Laughlin, G.\ 2006, preprint (astro-ph/0603376) 


\bibitem[Charbonneau et al.(2006)]{2006ApJ...636..445C} Charbonneau, D., et 
al.\ 2006, \apj, 636, 445 


\bibitem[Chiang \& Lithwick(2005)]{2005ApJ...628..520C} Chiang, E.~I., \& 
Lithwick, Y.\ 2005, \apj, 628, 520 


\bibitem[Cresswell \& Nelson(2006)]{2006A&A...450..833C} Cresswell, P., \& 
Nelson, R.~P.\ 2006, \aap, 450, 833 

\bibitem[Deming et al.(2005)]{2005Natur.434..740D} Deming, D., Seager, S., 
 Richardson, L.~J., \& Harrington, J.\ 2005, \nat, 434, 740 

\bibitem[Ford(2005)]{2005AJ....129.1706F} Ford, E.~B.\ 2005, \aj, 129, 1706 

\bibitem[Ford(2006)]{2006ApJ...642..505F} Ford, E.~B.\ 2006, \apj, 642, 505 

\bibitem[Ford \& Chiang(2006)]{FordChiang06} Ford, E.~B., \& Chiang, E.~I. 2006, in prep.

%\bibitem[Ford \& Gaudi(2006)]{FordGaudi06} Ford, E.~B., \& Gaudi, B.~S. 2006, submitted to ApJL

\bibitem[Ford \& Rasio(2006)]{2006ApJ...638L..45F} Ford, E.~B., \& Rasio, 
F.~A.\ 2006, \apjl, 638, L45 

\bibitem[Gaudi(2003)]{2003astro.ph..7280G} Gaudi, B.~S.\ 2003, preprint (astro-ph/0307280)

\bibitem[Gaudi \& Winn(2006)]{2006astro.ph..8071G} Gaudi, B.~S., \& Winn, 
J.~N.\ 2006, preprint (astro-ph/0608071)

\bibitem[Go{\'z}dziewski \& Konacki(2006)]{2006ApJ...647..573G} 
Go{\'z}dziewski, K., \& Konacki, M.\ 2006, \apj, 647, 573 

\bibitem[Gregory(2005)]{2005ApJ...631.1198G} Gregory, P.~C.\ 2005, \apj, 631, 1198 

% \bibitem[Harrington et al. (2006)]{2006HarringtonInPrep} Harrington, J., Luszcz, S.H., Deming, D., Seager, S., Richardson, L.J. 2006, in preparation.

\bibitem[Holman \& Murray(2005)]{2005Sci...307.1288H} Holman, M.~J., \& 
Murray, N.~W.\ 2005, Science, 307, 1288 

\bibitem[Holman et al.(2006)]{2006astro.ph..7571H} Holman, M.~J., et al.\ 
2006, preprint (astro-ph/0607571)

\bibitem[Ji et al.(2005)]{2005ApJ...631.1191J} Ji, J., Liu, L., Kinoshita, 
H., \& Li, G.\ 2005, \apj, 631, 1191 

\bibitem[Knutson et al.(2006)]{2006astro.ph..3542K} Knutson, H., 
Charbonneau, D., Noyes, R.~W., Brown, T.~M., \& Gilliland, R.~L.\ 2006, 
preprint (astro-ph/0603542)


\bibitem[Kortenkamp et al.(2004)]{2004Icar..167..347K} Kortenkamp, S.~J., 
Malhotra, R., \& Michtchenko, T.\ 2004, Icarus, 167, 347 


\bibitem[Laughlin \& Chambers(2002)]{2002AJ....124..592L} Laughlin, G., \& 
Chambers, J.~E.\ 2002, \aj, 124, 592 


% \bibitem[Marzari \& Scholl(2002)]{2002Icar..159..328M} Marzari, F., \& Scholl, H.\ 2002, Icarus, 159, 328 

\bibitem[]{} Michtchenko, T.A. \& Malhotra, R. 2004, Icarus, 168, 237.

\bibitem[McCullough et al.(2006)]{2006McCullough} McCullough, P.R. et al.\ 2006, ApJ, 648, 1228.

\bibitem[Morbidelli et al.(2005)]{2005Natur.435..462M} Morbidelli, A., 
Levison, H.~F., Tsiganis, K., \& Gomes, R.\ 2005, \nat, 435, 462 


\bibitem[Murray \& Dermott(2000)]{2000ssd..book.....M} Murray, C.~D., \& 
Dermott, S.~F.\ 2000, Solar System Dynamics, Cambridge, UK: Cambridge University Press.
%, by C.D.~Murray and S.F.~Dermott.~ ISBN 0521575974.~http://www.cambridge.org/us/catalogue/catalogue.asp?isbn=0521575974.~Cambridge, UK: Cambridge University Press, 2000.

\bibitem[Nesvorn{\'y} \& Dones(2002)]{2002Icar..160..271N} Nesvorn{\'y}, 
D., \& Dones, L.\ 2002, Icarus, 160, 271 

\bibitem[Nesvorn{\'y} et al.(2002)]{2002CeMDA..82..323N} Nesvorn{\'y}, D., Thomas, F., Ferraz-Mello, S., \& Morbidelli, A.\ 2002, Celestial Mechanics and Dynamical Astronomy, 82, 323 

\bibitem[ODonovan et al.(2006)]{2006ODonovanPreprint} O'Donovan, F.R. et 
al.\  2006, ApJL, in press

\bibitem[Rasio et al.(1996)]{1996ApJ...470.1187R} Rasio, F.~A., Tout, 
C.~A., Lubow, S.~H., \& Livio, M.\ 1996, \apj, 470, 1187 

\bibitem[Rasio \& Ford(1996)]{1996Sci...274..954R} Rasio, F.~A., \& Ford, 
E.~B.\ 1996, Science, 274, 954 

\bibitem[Rowe et al.(2006)]{2006ApJ...646.1241R} Rowe, J.~F., et al.\ 2006, 
\apj, 646, 1241 

\bibitem[Sato et al.(2005)]{2005ApJ...633..465S} Sato, B., et al.\ 2005, 
\apj, 633, 465

\bibitem[Schwarz et al.(2005)]{2005AsBio...5..579S} Schwarz, R., 
Pilat-Lohinger, E., Dvorak, R., {\'E}rdi, B., \& S{\'a}ndor, Z.\ 2005, 
Astrobiology, 5, 579 

\bibitem[Thommes(2005)]{2005ApJ...626.1033T} Thommes, E.~W.\ 2005, \apj, 
626, 1033 

%\bibitem[Winn \& Holman(2005)]{2005ApJ...628L.159W} Winn, J.~N., \& Holman, M.~J.\ 2005, \apjl, 628, L159 

\bibitem[Winn et al.(2005)]{2005ApJ...631.1215W} Winn, J.~N., et al.\ 2005, 
\apj, 631, 1215 

\bibitem[Wright(2005)]{2005PASP..117..657W} Wright, J.~T.\ 2005, \pasp, 
117, 657 

\bibitem[Wu \& Murray(2003)]{2003ApJ...589..605W} Wu, Y., \& Murray, N.\ 
2003, \apj, 589, 605 

\end{thebibliography}
\end{document}